\documentstyle[aps,prl,psfig,epsfig,floats,twocolumn]{revtex}

\begin{document}
\wideabs{
\title{Electron Glass in Ultrathin Granular Al Films at Low Temperatures}

\draft

\author{E. Bielejec and Wenhao Wu}
\address{Department of Physics and Astronomy, University of Rochester,}
\address{Rochester, New York 14627}

\date{\today}

\maketitle

\begin{abstract}

Quench-condensed granular Al films, with normal-state sheet
resistance close to 10 k$\Omega$/$\Box$, display strong hysteresis
and ultraslow, non-exponential relaxation in the resistance when
temperature is varied below 300 mK. The hysteresis is nonlinear
and can be suppressed by a dc bias voltage. The relaxation time
does not obey the Arrhenius form, indicating the existence of a
broad distribution of low energy barriers. Furthermore, large
resistance fluctuations, having a $1/f$-type power spectrum with a
low-frequency cut-off, are observed at low temperatures. With
decreasing temperature, the amplitude of the fluctuation increases
and the cut-off frequency decreases. These observations combine to
provide a coherent picture that there exists a new glassy electron
state in ultrathin granular Al films, with a growing correlation
length at low temperatures.

\end{abstract}

}

The electron glass was first predicted to exist in disordered
interacting systems nearly two decades ago \cite{Davies}. Such
nonergodic behavior is very interesting because one normally
expects electron systems to relax rather rapidly. Over the years,
a number of studies have been reported in which electrons display
glassy dynamics that are often associated with non-exponential
relaxation extending over many decades in time, such as the
field-effect conductance measurements in compensated GaAs
\cite{Monroe}, amorphous indium-oxide films \cite{Ovadyahu}, and
ultrathin Bi/Ge and Pb/Ge films \cite{Martinez}. The glassy
behavior is believed to arise from the electron-electron ({\em
e-e}) interactions and the Coulomb gap \cite{Pollak,Efros,Yu}.
Recently, the electron glass has received renewed interest
\cite{Orignac} as the subject of {\em e-e} interactions has become
a central topic in understanding the metal-insulator transition in
two-dimensions \cite{Simonian}. However, the precise role that the
Coulomb gap plays in the observed glassy behavior is not clear
since there is no simultaneous measurement of the conductance
relaxation and the single particle density of states.  In
addition, there is no direct experimental effort to probe the
correlation length in the glassy phase.

In this Letter we report glassy behavior in the normal state of
quench-condensed weakly insulating granular Al films of sheet
resistance, $R_{\Box}$, of about 10 k$\Omega$/$\Box$ at 300 mK. We
measure the relaxation of $R_{\Box}$ after the temperature was
varied. We have focused on weakly insulating films because
$R_{\Box}$ on the order of 10 k$\Omega$/$\Box$ is easy to measure
using sensitive ac lock-in techniques, which turn out to be
crucial in measuring the resistance fluctuations described below.
We observed that, below 300 mK, the resistance was strongly
hysteretic and displayed ultraslow, non-exponential relaxation as
the temperature was varied. We have also observed strong nonlinear
behavior in the hysteretic regime. What was unique to our work was
the first observation in quench-condensed metal films of large
resistance fluctuations below 100 mK. We argue that these
observations indicate the existence of a glassy electron state in
ultrathin granular Al films with a growing correlation length at
low temperatures.

Our Al films were quench-condensed onto glass substrates using
99.999$\%$ purity Al sources in UHV conditions inside a dilution
refrigerator, with the substrates being held near 20 K during
evaporation. The film thickness was near 25 {\AA}. After warming
up to room temperature, the films showed a granular morphology
with a typical grain size of about 300 {\AA}, as seen in scanning
force microscopy studies. The films had a multi-lead pattern with
an area of 3$\times$3 mm$^{2}$ between the neighboring leads. An
analog lock-in amplifier, operating at 27 Hz with a time constant
of 3 seconds, was used to measure the four-probe ac resistance.
The ac probe current was fixed at 1 nA, producing an ac voltage of
about 10 $\mu$V across a film section. For nonlinear studies
described below, an additional dc bias voltage was applied. All
the data described below were measured in the normal state, with
superconductivity being suppressed by a magnetic field above the
spin-paramagnetic limit \cite{Tedrow,Wu} of about 48 kG. To date,
we have performed detailed studies on films of $R_{\Box}$ $\sim$
10 k$\Omega$/$\Box$. Such films appeared to be very uniform, with
$R_{\Box}$ varying less than 5$\%$ among the various sections of
the multi-lead pattern. These films are far above the percolation
threshold since $R_{\Box}$ scales with inverse film thickness.

\begin{figure}[t]
\centerline{\epsfig{file=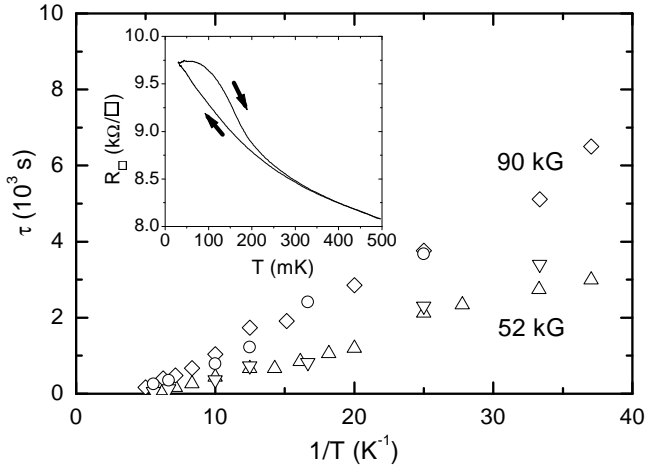,width=8.5cm}} \caption{Main
figure shows time constant $\tau$ vs. $1/T$ for various applied
fields: perpendicular 52 kG (up triangles), parallel 52 kG (down
triangles), perpendicular 90 kG (circles), and parallel 90 kG
(diamonds). Inset shows a typical resistance hysteresis, with the
arrows indicating the direction of the temperature sweeps.}
\label{Figure 1}
\end{figure}

In the inset to Fig. 1, we show a typical resistance hysteresis
measured on one Al film as temperature was cycled between 500 and
30 mK, in the high-field normal state with a field, $H_{\perp}$
$=$ 52 kG, applied perpendicular to the film plane. Throughout all
our experiments, the rates of cooling and heating were kept
constant on a logarithmic temperature scale, and the cooling and
the heating cycles each took 1.5 hr. We can exclude temperature
lag or heating as the cause of the hysteresis, because we have not
observed any hystersis in Li films of $R_{\Box}$ $\sim$ 10
k$\Omega$/$\Box$ and Be film of $R_{\Box}$ $=$ 3 $\sim$ 500
k$\Omega$/$\Box$, both quench-condensed on the same glass
substrates. For relaxation studies, the films were first cooled
from 500 mK to the desired temperature, $T$, using the above
logarithmic rate, at which $R_{\Box}$ was measured as a function
of time. In Fig. 2, we show the $R_{\Box}$ vs. time curves for a
number of fixed $T$ in a field of $H_{\perp}$ $=$ 52 kG. We found
that the curves at higher $T$, such as 100 mK, could be fit very
well to a stretched-exponential form,
$[R_{\Box}(\infty)-R_{\Box}(t)]/[R_{\Box}(\infty)-R_{\Box}(0)] =
exp[-(t/ \tau)^{\gamma}]$, over three decades in time. With
decreasing $T$, the time constant $\tau$ increased sharply. The
exponent $\gamma$ scattered between 0.6 and 0.8 at low $T$ without
a clear trend. We note that below 100 mK, the large resistance
fluctuations seen in Fig. 2 made it difficult to accurately
determine $\tau$ and $\gamma$. In the main part of Fig. 1, we plot
$\tau$ vs. $1/T$ for two field values in both perpendicular and
parallel field orientations. Data in Fig.1 appear to fall into two
groups: one at 52 kG and the other at 90 kG, suggesting that
$\tau$ was larger in higher fields and was insensitive to field
orientation. The almost linear dependence of $\tau$ on $1/T$ in
Fig. 1 indicates that, with decreasing $T$, $\tau$ increased much
slower than the Arrhenius law, $\tau(T) = \tau_{0}exp(E_{a}/T)$,
where $E_{a}$ is known as the activation energy describing the
typical energy barrier to relaxation \cite{Binder}. Such
non-Arrhenius behavior can be explained only if there exists a
very broad distribution of low-energy barriers. Relaxation occurs
over lower energy barriers with decreasing $T$, leading to a
relaxation time that increases slower than the Arrhenius law.

\begin{figure}[t]
\centerline{\epsfig{file=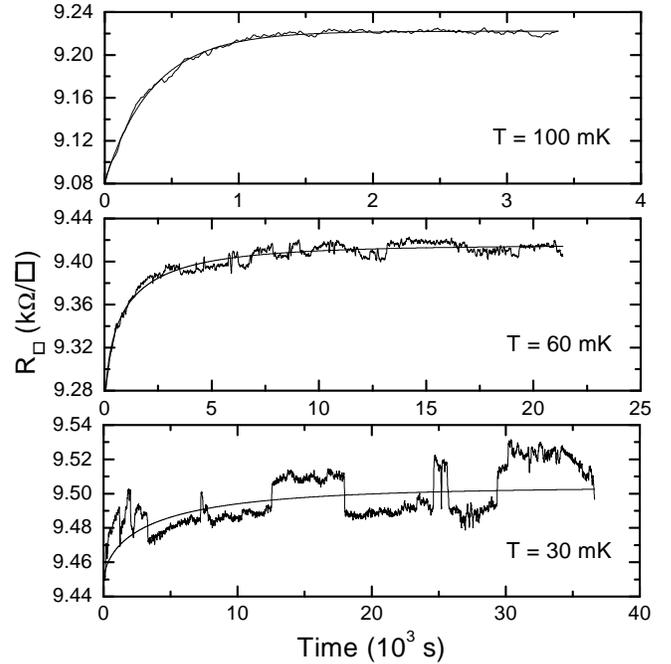,width=8.5cm}} \caption{Time
traces of film sheet resistance measured at temperatures as
labeled on the graph, in a field of $H_{\perp} =$ 52 kG. Smooth
lines are fits to a stretched-exponential form.} \label{Figure 2}
\end{figure}

The above conclusion is also supported by the nonlinear effects in
the hysteresis. For nonlinear studies, we have measured $R_{\Box}$
in the presence of a dc bias voltage, $V_{bias}$, across the
films, and have observed that the hysteresis can be suppressed if
$V_{bias}$ is large enough. The experimentally relevant $V_{bias}$
was of the order of 0.5 mV. It leads to a bias current of 0.05
$\mu$A and a Joule heating of 2.5$\times$10$^{-11}$ W. We believe
that the nonlinear effects were not caused by heating. Consider
the hysteresis shown in the inset to Fig. 1, we observed that,
between 100 mK and 200 mK, the heating curve shifted downward and
the cooling curve shifted upward with increasing $V_{bias}$. It is
obvious that had heating by the bias current been significant the
cooling curve should have also shifted downward with increasing
$V_{bias}$, since the resistance of weakly insulating films should
decrease with increasing $T$. In Fig. 3(a), we plot the width of
the hysteresis, $\Delta R$, versus $T$ for various $V_{bias}$
values. Such data are obtained by subtracting the cooling curve
from the heating curve in a thermal cycle such as the one shown in
the inset to Fig. 1. It is quite revealing in that the bias
voltage does not uniformly suppress the hysteresis across the
entire range of $T$. Instead, a small $V_{bias}$ suppresses the
low-$T$ components of the hysteresis only. With increasing
$V_{bias}$, progressively higher-$T$ components of the hysteresis
are suppressed. We believe that relaxation processes over
low-energy barriers are suppressed by the bias voltage in the same
way the low-$T$ components of the hysteresis are suppressed.
Nevertheless, we have yet to develop quantitative methods to
analyze the data in Fig. 3(a) to obtain the distribution of
low-energy barriers. We note that the falling off of $\Delta R$ at
low $T$ in the $V_{bias}$ $=$ 0 curve in Fig. 3(a) was an
experimental artifact because we did not wait at 30 mK for the
system to relax after cooling down from 500 mK. Had we waited long
enough before heating up towards 500 mK, $\Delta R$ would have
been the largest at 30 mK for the $V_{bias}$ $=$ 0 curve.

\begin{figure}[t]
\centerline{\epsfig{file=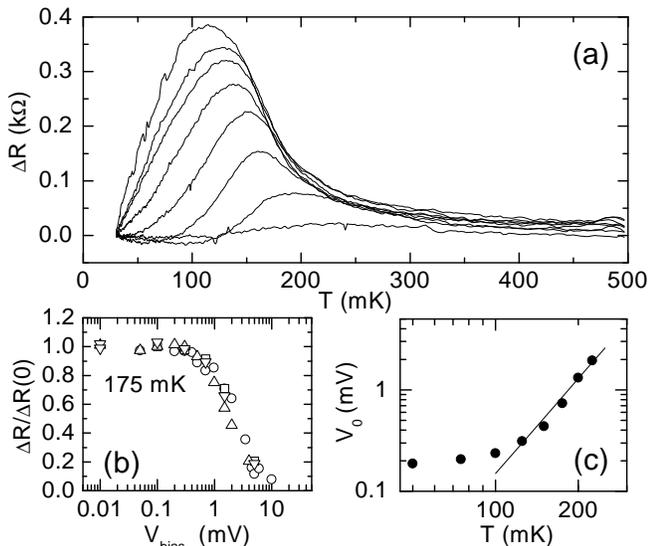,width=8.5cm}} \caption{(a)
shows the hysteresis width as a function of temperature for
$H_{\perp} =$ 52 kG. Curves from top to bottom correspond to
$V_{bias} =$ 0, 0.3, 0.4, 0.5, 0.98, 1.5 3.5, and 10 mV,
respectively. (b) shows the scaled hysteresis width vs. $V_{bias}$
at 175 mK. There are four types of symbols, corresponding to
applied fields of 52 kG or 90 kG, in either perpendicular or
parallel orientations. (c) shows a log-log plot of the temperature
dependence of the threshold voltage. The solid line is a power-law
fit with a power of 3.1 $\pm$ 0.3.} \label{Figure 3}
\end{figure}

If we make a vertical cut at a fixed $T$, such as 175 mK, in Fig.
3(a), we obtain $\Delta R$ as a function of $V_{bias}$. We have
observed that the absolute value of $\Delta R$ varied with the
strength and the orientation of the magnetic field. However, if we
scale $\Delta R$ by its value in the $V_{bias}$ $=$ 0 limit,
$\Delta R(0)$, the resulting scaled width, $\Delta R$/$\Delta
R(0)$, appears to be independent of the strength and the
orientation of the field. This is demonstrated by the good data
collapse in Fig. 3(b) in which we plot four sets of such scaled
data at 175 mK as a function of $V_{bias}$ for two field values in
both perpendicular and parallel orientations. In this semi-log
plot, $\Delta R$ falls off steeply for $V_{bias}$ $>$ 0.7 mV. This
suggests that there is a field-independent threshold bias, $V_{0}$
$\sim$ 0.7 mV, that suppresses the hysteresis. By making vertical
cuts at various temperatures in Fig. 3(a), we observe that $V_{0}$
decreases with decreasing $T$. This is consistent with the picture
that relaxation occurs over lower energy barriers at lower $T$; as
a result, the value of $V_{0}$ for suppressing the hysteresis is
also lowered. We plot $V_{0}$ vs. $T$ in Fig. 3(c). The flattening
of the data below 100 mK is again due to the experimental artifact
discussed earlier. For the high-$T$ part of the data, $V_{0}$ can
be fit to a power law temperature dependence, $V_{0}$ $\sim$
$T^{3.1 \pm 0.3}$, as shown by the solid line in the inset to Fig.
3.

The results discussed above suggest the existence of a glassy
state in our quench-condensed ultrathin granular Al films. The low
temperature scale of our experiments excludes the possibility of
glassy structural relaxation, such as atomic diffusion between
metastable configurations similar to those observed in a-Si:H
\cite{Kakalios}. Tunneling studies have shown in such Al films
\cite{Wu2}, as well as in other films of similar sheet resistance
\cite{White}, that an anomaly in the density of state exists at
the Fermi energy. However, a true Coulomb gap has not been found
in such weakly insulating films. Thus it is not clear what role,
if any, the Coulomb gap plays in the observed glassy relaxation.
In addition, we did not find any hysteresis in quench-condensed Li
films of $R_{\Box}$ $\sim$ 10 k$\Omega$/$\Box$ and Be films of
$R_{\Box}$ $=$ 3 $\sim$ 500 k$\Omega$/$\Box$, even though a true
Coulomb gap has been observed in Be films \cite{Bielejec}. While
we were not able to investigate the morphology of the Li films,
which become unstable in air, scanning force microscopy studies of
the Be films warmed up to room temperature did not find any
granular structure down to 1 nm. Because of the morphological
difference, we propose that granularity plays an important role in
the glassy behavior of our Al films.

Consider a simple conduction model for granular metals in which an
electron moves from one neutral grain to another nearby one to
create a charge-anticharge pair \cite{Neugebauer}. The energy cost
for creating such a pair is 2$E_{c}$, where $E_{c} = e^{2}/2C$ is
the grain charging energy, with C being the capacitance of a
grain. $E_{c}$ can be significant if the grains are small, and can
vary strongly with location due to the randomness in grain sizes.
This results in a rough potential background for the charge
carriers. Transport is mediated via the ionization of such pairs
\cite{Abeles}, with the energy associated with ionization being
the Coulomb attraction between the pair, which, in two dimensions,
has a logarithmic form with a cutoff length \cite{Mooij}. This
model has motivated experiments searching for a finite-$T$
Kosterlitz-Thouless-Berezinskii charge unbinding transition in
arrays of Josephson junctions \cite{Mooij,Mooij2} and granular
films \cite{Liu}. However, in a recent theoretical study of
capacitively coupled grains, Granato and Kosterlitz \cite{Granato}
have found that finite-$T$ transitions are suppressed by disorder.
They found instead a charge glass with a correlation length that
diverges as temperature decreases to zero. They have also
predicted that nonlinear behavior sets in at a characteristic
voltage $V_{c} \sim T^{1+ n}$, with a thermal critical exponent
$n$ $\approx$  1.7 characterizing the $T$ $=$ 0 charge glass
transition. It is interesting to note the closeness of this
predicted power-law dependence with the power-law fit in Fig.
3(c). However, we caution that our fit has a limited range. Within
this picture of granular transport, we can qualitatively
understand the behavior of the films in an applied magnetic field
as shown in the main part of Fig. 1.  The field splits the singlet
states, effectively reducing the density of states.  As a result,
the relaxation time increases with increasing magnetic field,
however, it is insensitive to the orientation of the field.

\begin{figure}[t]
\centerline{\epsfig{file=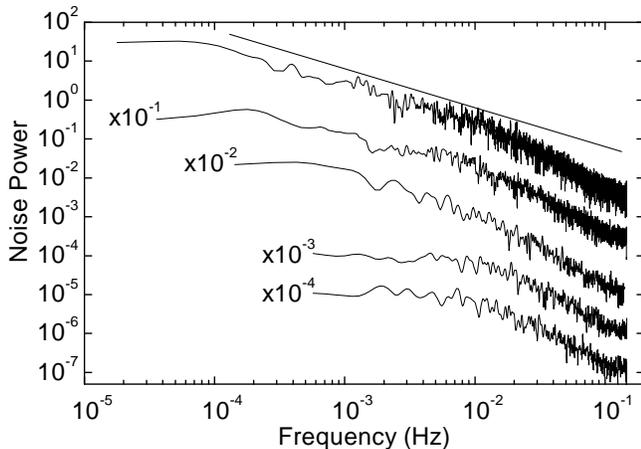,width=8.5cm}} \caption{Power
spectra of the resistance fluctuations are shown for 30, 60, 80,
100, and 120 mK, corresponding to curves from top to bottom,
respectively. Four of the five curves are shifted by factors
labeled next to the curves. The straight line shows the $1/f$-type
frequency dependence.} \label{Figure 4}
\end{figure}

We note that the data in Fig. 2 show drastic increases in the size
and decreases in the characteristic frequency of the resistance
fluctuations with decreasing $T$, as shown by the 60 mK and 30 mK
curves. This is a unique feature in our data that has not been
reported in similar quench-condensed films. We have analyzed such
noise data using a fast Fourier transform (FFT). We first obtain
the noise time traces by subtracting the stretched-exponential
fitting curves from the $R_{\Box}$ vs. time traces. The noise time
traces are then analyzed using FFT. Figure 4 shows a set of such
spectra obtained from data at a number of temperatures with
$H_{\perp}$ $=$ 52 kG, including the three traces shown in Fig. 2.
There are two distinguishing features in the spectra. The first is
that, the $T$ $=$ 30 mK spectrum, which has the broadest frequency
range, clearly has a $1/f$-type form \cite{Dutta} over a wide
range of frequency, indicating that fluctuations occur at all
frequencies. The second is that all the spectra in Fig. 4 level
off at low frequencies, indicating a low-frequency cut-off in the
spectra which clearly shifts to lower frequencies as $T$ is
lowered. The increase in the magnitude of the low-frequency
fluctuations and the decrease in the cut-off frequency with
decreasing $T$ in Fig. 4 indicate that the number of low-energy
states increases. We suggest that these two features arise from
the collective hopping of many correlated electrons.  The cut-off
frequency should be a measure of the characteristic frequency of
fluctuations occurring on the length scale of the correlation
length.  A decreasing cut-off frequency with decreasing $T$
indicates an increasing correlation length.  We also point out
that the power spectra for all the temperatures shown in Fig. 4
nearly follow a single curve for $f >$ 10$^{-2}$ Hz, with a
frequency dependence of $1/f^{1.72 \pm 0.08}$. Such behavior
deviates significantly from the $1/f$ behavior seen in the 30 mK
curve at low frequencies. Although the origin of such behavior is
unknown, it is possible that this nearly temperature independent
high-frequency feature is not related to collective electron
hopping. Rather, it is the emerging $1/f$ part of the spectra at
low temperatures that represents the appearance of collective
electron hopping. In general, we expect the following relation
between the cut-off frequency, $f_{c}$, the correlation length,
$\xi$, and $T$: $f_{c} \sim \xi^{-z} \sim T^{\nu z}$, where $\nu$
is the correlation length exponent and $z$ is the dynamical
critical exponent. Ideally, one measures the noise using a
sensitive bridge setup \cite{Scofield} in which the large and
non-fluctuating resistance background is eliminated, rather than
the simple four-terminal methods used in this study. In addition,
the fluctuations should be measured when the system is in a
steady-state, instead of in the initial relaxation process as
shown in Fig. 2.  Future experiments taking the above mentioned
concerns into account should provide an accurate determination of
$f_{c}$ and the critical exponents.

In conclusion, we have found strong evidence for a new glassy
electron state in quench-condensed ultrathin granular Al films,
with a growing correlation length at low temperatures. We
gratefully acknowledge numerous invaluable discussions with S.
Teitel, Y. Shapir, Y. Gao, and P. Adams. We thank S. Zorba and Y.
Gao who performed scanning force microscopy studies of the
quench-condensed Al and Be films.

%
%

%

%
\end{document}